# End-to-end deep learning for super-oscillatory subtraction imaging

Zhongwei Jin[a,b,*], Keyi Chen[a], Qiuyu Ren[a], Zhigang Dai[a], Ruoping Yao[a], Zhi Hong[b], Bin Fang[a,b], Fangzhou Shu[b,*] Shengtao Mei[a,], Yiping Lu[a,*]
[a]*College of Optical and Electronic Technology, China Jiliang University, Hangzhou 310018, China*
[b]*Centre for THz Research, China Jiliang University, Hangzhou 310018, China*
**jinzhongwei@cjlu.edu.cn*
**fzshu@cjlu.edu.cn*
**20a0402180@cjlu.edu.cn*

**Abstract:** Breaking the diffraction limit in optical imaging is crucial for resolving subwavelength details in a wide range of applications, where superoscillatory imaging and subtraction imaging are two common strategies for surpassing conventional resolution limits. We propose an end-to-end deep learning framework that integrates superoscillatory focusing and subtraction imaging into a single jointly-optimized vectorial Debye integral neural network pipeline, eliminating the traditional two-step acquisition and manual weighting process. With this end-to-end neural network, we further improve the focusing capability of the system to the sub-100-nm regime, enabling deep-subwavelength imaging resolution.

Extensive efforts have been made to enhance optical imaging resolution, with super-resolution techniques drawing focus for their ability to surpass the conventional diffraction limit (~250 nm) [1, 2]. Widely adopted far-field methods such as stimulated emission depletion microscopy (STED) [3, 4], structured illumination microscopy (SIM) [5, 6], and stochastic optical reconstruction microscopy (STORM) [7, 8] routinely achieve resolutions of several tens of nanometers, greatly advancing research in chemistry, biology, and medicine [9-11]. Meanwhile, the pursuit of higher resolution continues [12, 13]. Super-oscillatory (SO) imaging is one such label-free far-field technique, in which carefully engineered interference fields exhibit local oscillations faster than the diffraction limit, enabling far-field super-resolution [14-19]. Subtraction imaging (SI), by contrast, achieves resolution enhancement through a mechanism reminiscent of STED: two beams combine to form an effective point spread function (PSF) that exceeds the diffraction limit [20-22].

While discovering entirely new super-resolution principles is challenging, integrating established techniques offers a practical pathway for further improvement. For instance, introducing an SO subwavelength periodic field into SIM has been shown to boost resolution [23], and combining SI with fluorescence microscopy has given rise to fluorescence emission difference microscopy (FED) [24]. Consequently, we reason that merging SO focusing with SI could deliver enhanced resolution. In parallel, deep neural network has emerged as a versatile framework, showing great potential in scientific research [25, 26], including imaging tasks [27, 28]. Here, we present an end-to-end neural-network pipeline that jointly optimizes both SO focusing and SI. This approach not only overcoming the long-standing challenge of manually tuning the subtraction coefficient but also producing an equivalent PSF with a smaller full width at half maximum (FWHM).

In our previous work, we introduced the vectorial Debye integral neural network (VDINN) which can produce superoscillatory foci with very low sidelobes [16]. Previous studies have shown that the FWHM of most SO foci is typically above ~0.3 $\lambda_m$ ($\lambda_m = \lambda/n$, $n$ is the refractive index), as smaller FWHM values often entail prohibitively low energy efficiency in practice [14]. Our earlier results agreed with this trend: within a defined Field of View (FoV) ($d < 2.5$μm), we obtained a minimum SO focal spot of 0.33 $\lambda_m$ [16].

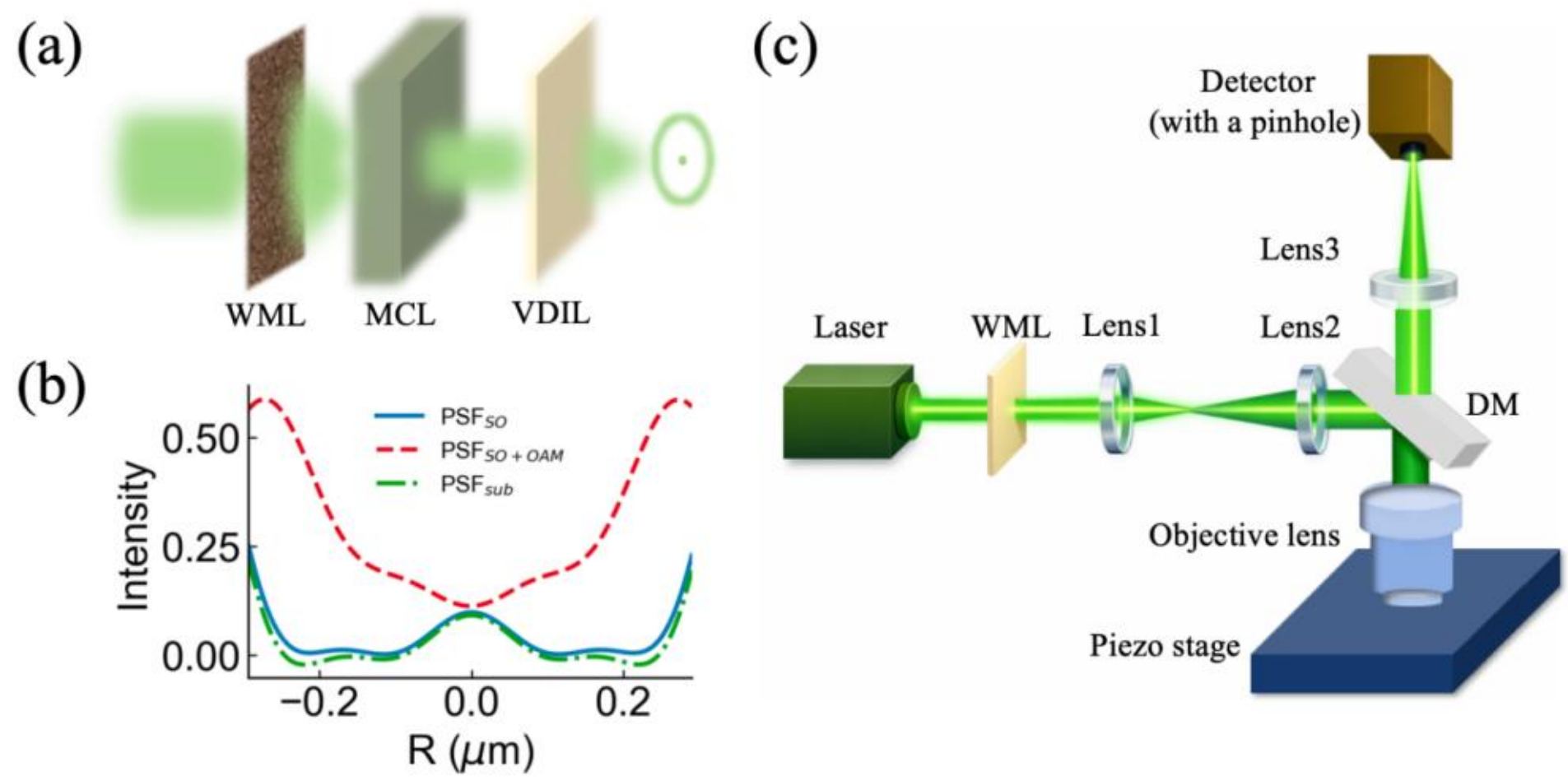

**Fig. 1.** VDINN scheme, calculated PSFs and optical setup. (a). Three parts of VDINN framework. (b). Three 1D PSFs from a trained VDINN and its SI by simply adding vortex phase. (c). Schematic of the simulated confocal laser scanning microscopy imaging system.

Figure 1(a) illustrates the architecture of our VDINN, comprising, from left to right, a Wavefront Modulation Layer (WML), Multiple Convolutional Layer (MCL), and a Vectorial Debye Integral Layer (VDIL). Loss function of the focusing field is defined as follows:

$$L_{\text{total}} = \frac{1}{N}\sum_{r}\gamma(r)\left(\sqrt{E_{\text{x}}^{2}+E_{\text{y}}^{2}+E_{\text{z}}^{2}} - T\right)^{2}$$

$$\gamma(r) = \begin{cases} 1 & r < kR_{\max} \\ 0 & r \geq kR_{\max} \end{cases} \qquad T[i,j] = \begin{cases} 1, \text{ if } (i,j) = (512,512) \\ 0, \text{ if } (i,j) = \text{ otherwise} \end{cases} \qquad (1)$$

Here, $R_{max}$ represents the maximum radius of the entire focal plane, and $k$ defines its boundary between the inner and outer regions (We divide the focusing field into two regions). $N$ is the total number of pixels in the calculation grid. $T$ denotes the desired focusing field distribution as a (1024, 1024) tensor (covering a full physical FoV of 3μm×3μm), which serves as the ground truth. The overall structure is identical to that in our previous work [16]. In this study, we use 532-nm laser source and set the objective numerical aperture to $NA = 1.4$ (n = 1.52). The simulation aperture on the focal plane is defined as $R_{max}$ = 1.5 μm, with a corresponding pixel size (sampling interval) of approximately 2.93 nm, and the focal-plane region parameter is set to $k = 0.125$, so that VDINN computes the loss only over the central one-eighth of the predicted focal field. The smaller prediction window and more localized loss constraint are expected to yield superoscillatory foci with reduced FWHM. The VDINN was implemented using PyTorch. The PSF optimization was performed by training to minimize the total loss defined in Eq. (1). We utilized the Adam optimizer with an initial learning rate of $1.0\times10^{-3}$, which was dynamically adjusted using a Cosine Annealing scheduler. The training was conducted for 600 epochs with a batch size of 1. To prevent over-fitting and encourage spatial smoothness of the optimized mask, a weight decay of $1.0\times10^{-5}$ was applied. The entire process was accelerated by an NVIDIA GeForce RTX 4090 GPU, and convergence was typically achieved within 5 minutes.

In Figure 1(b), the blue solid curve corresponds to the superoscillatory $PSF_{SO}$ obtained by training, with an FWHM of 0.315 $\lambda_m$ (~110 nm). Statistical surveys of prior studies indicate that an FWHM of ~0.3 $\lambda_m$ represents the practical upper limit for experimentally viable superoscillatory focusing. Although higher resolutions are theoretically possible, they carry excessive experimental penalties. Following our hypothesis, combining SI with SO focusing could further improve resolution. SI involves, in addition to bright-field imaging, using a donut-shaped beam to construct a dark-field image, where the vortex beam is the most straightforward choice. Incorporating a vortex phase may not be the globally optimal choice—for instance, an azimuthally polarized beam potentially provides a finer dark-field energy distribution—adding a vortex phase to the WML is also straightforward experimentally and the specific selection of the dark-field beam does not affect the overall conclusions of our work. Experiment can be seamlessly achieved by inserting a q-plate or directly encoding the vortex phase onto the existing Wavefront Modulating Layer. Such a common-path strategy significantly minimizes systematic complexity and reduces the sensitivity to mechanical misalignment during beam switching. Consequently, we added a first-order vortex phase to the pre-trained wavefront-modulation mask. With the trained VDINN, we directly obtained the corresponding $PSF_{SO+OAM}$ (red dashed curve in Fig. 1b). Comparing $PSF_{SO}$ and $PSF_{SO+OAM}$, we observed a broader dark-field FWHM for the latter. This required choosing a smaller weighting factor $\gamma$, to achieve a meaningful subtraction-imaging effect as follows [20-22]:

$$PSF_{sub} = PSF_{SO} - \gamma PSF_{SO+OAM} \qquad (2)$$

Conventionally, we aim for the smallest FWHM possible, provided the negative sidelobes remain at an acceptable level. We finally selected $\gamma$=0.006, yielding a subtraction $PSF_{sub}$ (green dashed curve in Fig. 1b) with an FWHM of 0.267 $\lambda_m$, corresponding to an ideal resolution of 93 nm.

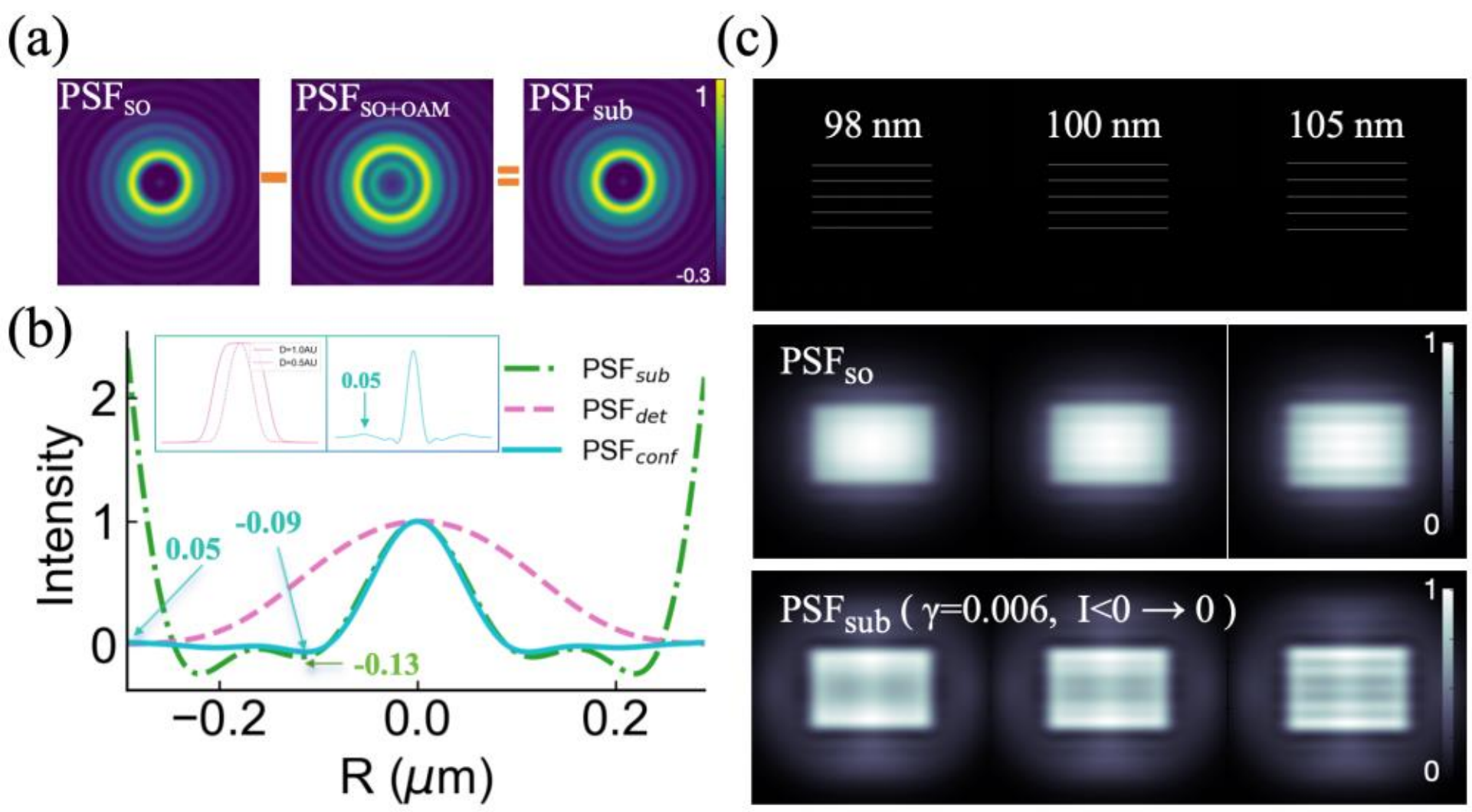

**Fig. 2.** Confocal imaging of simple combination of SO focusing and SI. (a). 2D PSFs subtracting process. (b). Total $PSF_{conf}$ of the confocal imaging system from pinhole filter's $PSF_{det}$ and synthetic $PSF_{sub}$ of SI. Inset left: $PSF_{det}$ with different pinhole size (dashed line: 0.5 AU, solid line: 1.0 AU). Inset right: $PSF_{conf}$ displayed over a larger spatial range. (c). Imaging simulation test. " $PSF_{SO}$" and "$PSF_{sub}$" imply the illumination field.

To verify the practical imaging performance of the aforementioned $PSF_{sub}$ within the entire confocal microscopy system, we still carry out imaging simulations under the confocal microscopy system as shown in Figure 1(c). In all subsequent simulation calculations, the center wavelength of the fluorescence emission signal is assumed to be 1.05 times the excitation wavelength. The effective PSF of the entire confocal imaging system is defined as $PSF_{conf} = PSF_{ext} \times PSF_{det}$, where $PSF_{ext}$ corresponds to the aforementioned $PSF_{SO}$ and $PSF_{SO+OAM}$, $PSF_{det}$ = convolution($PSF_{emi}$ , D), where $PSF_{emi}$ represents the PSF of the fluorescence signal as it passes through the objective lens, and D denotes the pinhole aperture in front of the detector [17, 18, 21, 22]. Since the final image is obtained by subtracting two separate acquisitions, the resulting effective PSF for a complete subtraction cycle is expressed as $PSF_{conf} = PSF_{sub} \times PSF_{det}$. Figure 2(a) shows the 2D PSFs for bright- and dark-field illumination and the resulting $PSF_{sub}$ obtained by their subtraction. This $PSF_{sub}$ represents only a reference, rather than the true $PSF_{sub}$ of the final post-processed image, since the subtraction inevitably introduces negative values that correspond to lost information. As SO foci are typically accompanied by strong sidelobes, using a confocal pinhole before the detector to suppress them is a common approach. Figure 2(b) presents the combined SO subtraction $PSF_{sub}$ after passing through a confocal pinhole of 0.5 Airy unit (AU) [17,18], yielding the effective $PSF_{conf}$. In the left part of the inset of Figure 2(b), the effect of different pinhole sizes on $PSF_{det}$ is presented. The solid curve corresponds to $PSF_{det}$ with a pinhole radius of 1.0 AU, whereas the dashed curve corresponds to a 0.5 AU pinhole, which is consistent with the $PSF_{det}$ profile shown in the main panel of Figure 2(b). It is evident that a 0.5 AU pinhole is sufficiently compact to effectively suppress the prominent sidelobes of $PSF_{sub}$. The peak intensity of the largest sidelobe in $PSF_{sub}$ is reduced to approximately 5% of the main lobe peak as shown in the right part of the inset. In contrast, employing a 1.0 AU pinhole would inevitably preserve a substantial fraction of the sidelobe energy in $PSF_{sub}$, which would consequently degrade the subsequent imaging performance. Simulations show that the pinhole effectively removes high-energy sidelobes outside the FoV, and simultaneously the relative normalized negative dip depth adjacent to the main lobe is reduced from 13 % to 9 % of the peak value. Although the negative values are suppressed, we can still observe information loss in the middle part of the grating in the third-row subfigure of Figure 2(c).

To evaluate the proposed method, we conducted an imaging simulation. In Figure. 2(c), the first row shows gratings to be imaged, with periods of 98 nm, 100 nm, and 105 nm. The second row shows the simulated bright-field imaging results when the SO focus generated by VDINN was used as the illumination in the confocal system. The effective PSF of the overall imaging system is expressed as $PSF_{2nd_row} = PSF_{SO} \times PSF_{det}$, the images were simulated by convolving the target grating objects with the composite system PSF (Similarly, the simulation results in the third row were obtained in the same manner). The results indicate a resolving capability between 100 nm and 105 nm, consistent with the empirical rule that imaging resolution slightly outperforms the FWHM of the PSF. The third row shows the SI results, obtained by subtracting 0.006 times the corresponding dark-field pattern generated with $PSF_{SO+OAM}$ from the brightfield image, with all negative values set to zero. Throughout our simulation, we do not directly modify any negative values within the effective $PSF_{sub}$. Instead, we only apply a zero-thresholding operation (clipping) to any negative intensities resulting from the final subtraction between the bright-field and dark-field images. For the 105-nm stripes, the combined SO subtraction-imaging (SOSI) approach produces patterns with improved contrast relative to SO alone. However, for the 100 nm and 98 nm cases, the performance degrades compared with pure SO imaging. This observation does not match the nominal FWHM = 93 nm of the synthetic $PSF_{sub}$. A likely reason is that the large negative dips in $PSF_{sub}$ lead to excessive information loss during post-processing. Supporting this, in the 100-nm pattern the central stripes appear dimmer than the lateral ones, causing the reconstructed image to lose essential structural information. Accordingly, the 5% maximum sidelobe intensity leads to noticeable halo-like artifacts around the resolved patterns, as seen in the simulation images. These results indicate that simply adding a vortex phase to the SO mask does not directly and effectively merge SO focusing with SI. The subtraction coefficient for combining bright- and dark-field images remains a subjective and sensitive hyperparameter [21, 22], which hinders robust integration of the two techniques.

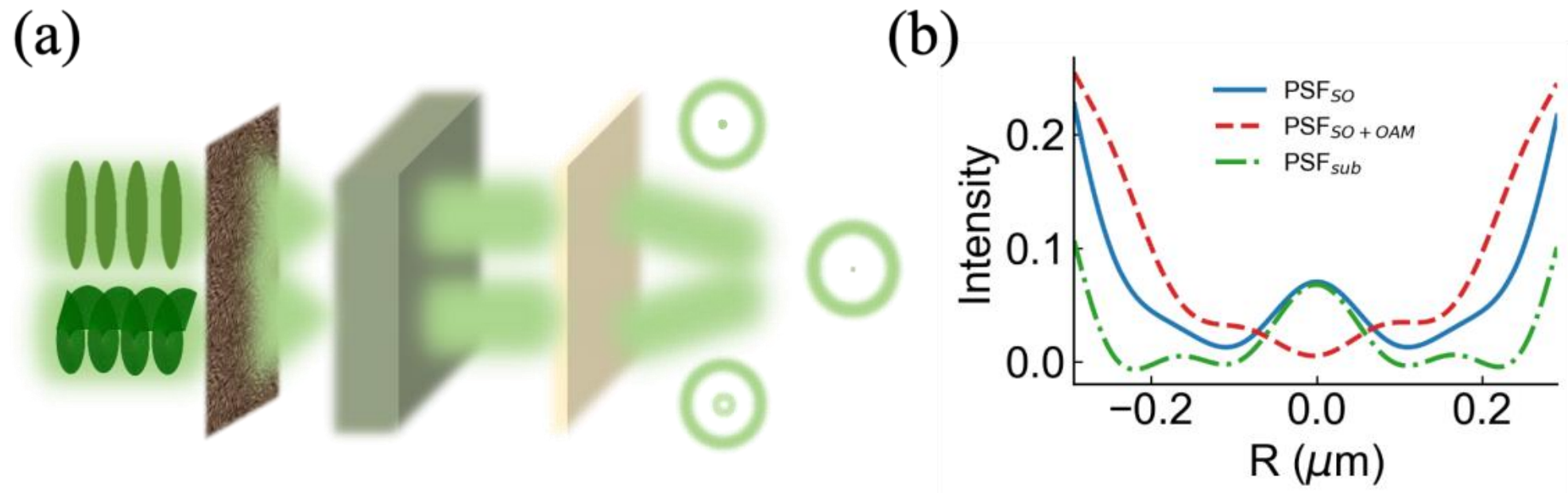

**Fig. 3.** End-to-end VDINN scheme for SOSI and its corresponding PSFs analysis. (a). End-to-end SOSI-VDINN configuration incorporates both plane-wave and vortex-beam inputs, and its output prediction is the result of subtracting the dark-field from the bright-field image. (b). Three 1D PSFs from a trained SOSI-VDINN.

To more effectively integrate SO and SI for stronger super-resolution capability, we further modified the VDINN training scheme by incorporating the two-step subtraction-imaging process within a single network. For the same wavefront-modulation mask, we simultaneously launch a plane wave and a first-order vortex beam, and subtract their focal-plane fields in the network output using a learnable parameter $\gamma$. In this way, VDINN performs end-to-end training of SOSI (Fig. 3(a)). Owing to the intrinsic nature of the MSE loss, values in the region surrounding the focal spot within the loss-calculation window are strongly suppressed. This not only removes high-energy sidelobes that degrade imaging but also inhibits negative dips that undermine SI. Moreover, $\gamma$ becomes a learnable hyperparameter in VDINN, rather than a subjectively tuned constant. Its criterion is the minimization of the global loss function. After training, the synthetic PSF$_{sub}$ (green dashed curve) is obtained (Fig. 3(b), with an FWHM of 0.284 $\lambda_m$, which is about 99 nm), together with the brightfield SO focus from the plane wave (blue solid curve, with an FWHM of 0.318 λm) and the dark-field distribution from the vortex beam (red dashed curve) through the same mask. Compared with Figure 1(b), both PSF$_{SO}$ and PSF$_{SO+OAM}$ exhibit significant changes in amplitude and shape, particularly the latter. These adjustments reduce the FWHM of the PSF$_{sub}$ main lobe and lower the depth of its nearest negative dip.

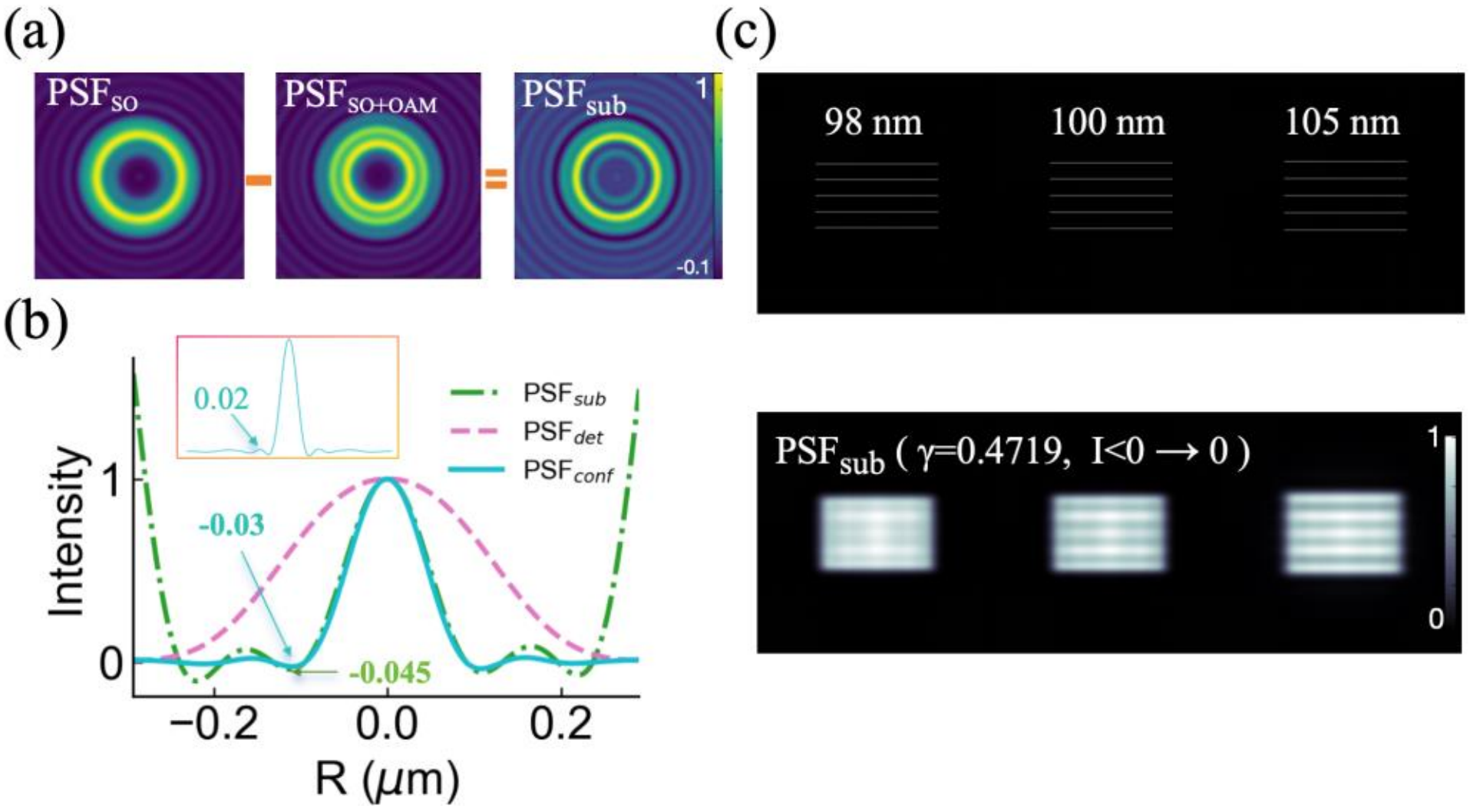

**Fig. 4.** Confocal imaging validation of end-to-end trained SOSI-VDINN. (a). 2D PSFs subtracting process after training. (b). Total PSF$_{conf}$ of the confocal imaging system from pinhole filter's PSF$_{det}$ and synthetic PSF$_{sub}$ of SI. Inset: PSFconf displayed over a larger spatial range. (c). Imaging simulation test. Imaging objects with different periods (1$^{st}$ row), SOSI-VDINN imaging test (2$^{nd}$ row).

The 2D subtraction composite PSF in Figure 4(a) further shows that the periphery of $PSF_{sub}$ carries more pronounced ripples than in Figure 2(a), likely the trade-off for reducing the nearest sidelobe's negative depth; with the pinhole in place, however, this trade-off has no impact on the final imaging. As shown in Figure 4(b), after the pinhole the relative normalized depth of the nearest negative dip is reduced from 4.5% to 3%, far below that in the simple vortex-phase-addition scheme. This substantial dip-depth reduction directly translates to higher system resolution. As illustrated in the inset of Figure 4(b), the peak intensity of the largest sidelobe has been optimized to 2% of the main lobe peak, which is significantly superior to the 5% level achieved through manual selection of $\gamma$ in Figure 2(b). In addition, from the energy distribution outside the main lobe, the profile obtained via end-to-end training is noticeably smoother and closer to zero compared to that obtained by manual tuning. This enhanced sidelobe suppression is explicitly verified by the imaging simulation results in Figure 4(c). Compared to Figure 2(c), the halo-like noise surrounding the resolved structures is noticeably weaker in Figure 4(c), a phenomenon that is also prominently reflected in the subsequent results of Figure 5.

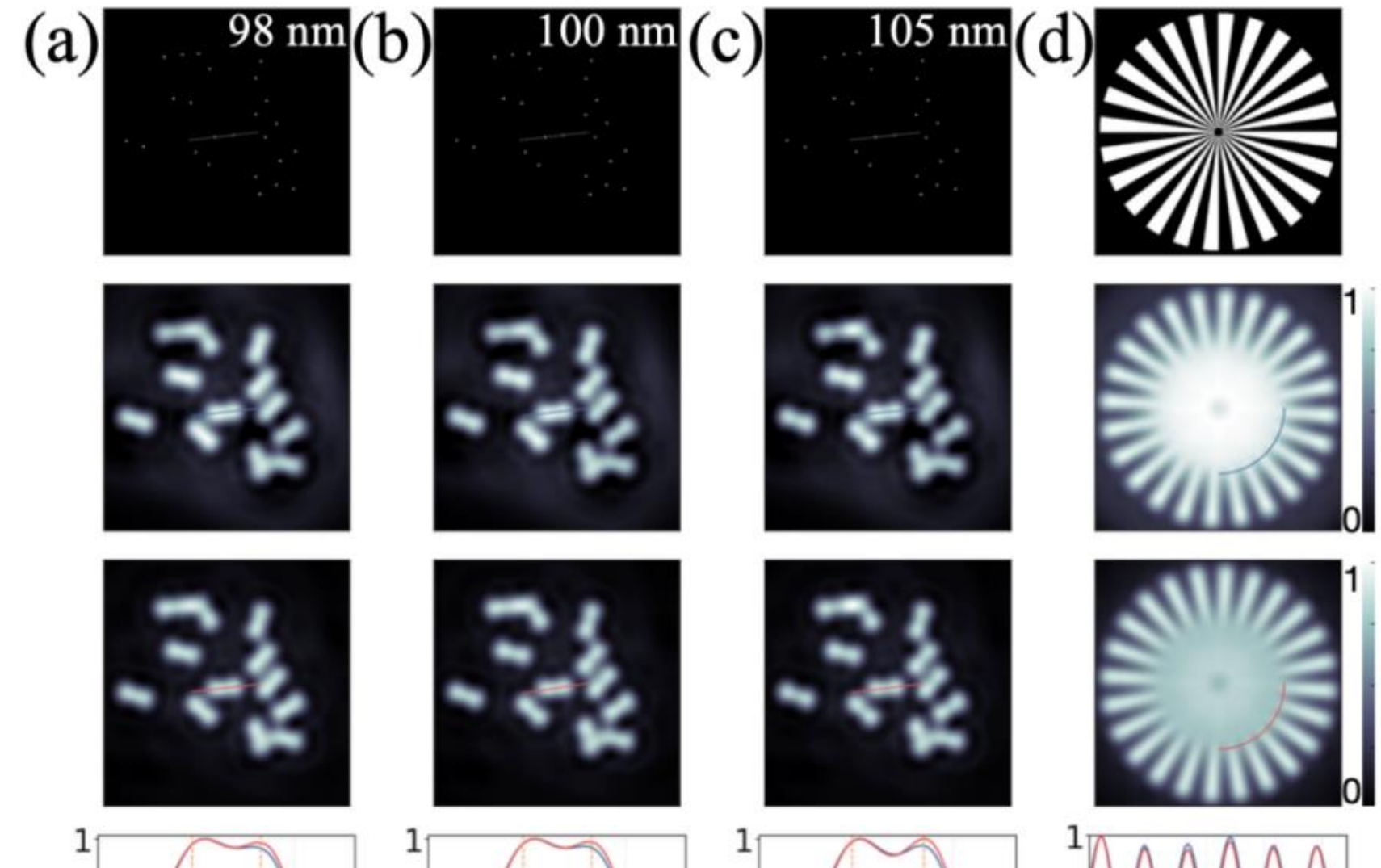

**Fig. 5**. Tests of two-point targets and Siemens star. (a) Imaging 98nm-seperated two-point targets. (b) Imaging 100nm-seperated two-point targets. (c) Imaging 105nm-seperated two-point targets. (d). Imaging of Siemens star (The red and blue arcs have identical radii).

To further validate the superiority of the end-to-end trained SOSI framework, tests of two-point resolution targets and a Siemens star pattern are shown in Figure 5. Figures 5(a)-(c) present the results for two-point targets with separations of 98 nm, 100 nm, and 105 nm, respectively, while Figure 5(d) shows the results for the Siemens star. The first row displays the original ground-truth patterns, which share consistent FoV and scale. The second row shows the imaging results obtained through the conventional SOSI method with manual parameter selection (Type 1), as described in Figure 2. The third row presents the results generated by the proposed end-to-end trained SOSI framework (Type 2), detailed in Figure 4. A visual comparison of these results reveals that the image clarity and contrast in Type 2 significantly outperform those in Type 1. Specifically, the Type 1 images exhibit noticeable halo-like artifacts and background noise surrounding the reconstructed structures, which stems from the stronger sidelobes in the $PSF_{conf}$ analyzed in Figure 2(b). The fourth row plots the intensity cross-sections along the solid lines indicated in the subplots of the second and third rows. In all test cases, the end-to-end SOSI demonstrates a superior ability to resolve the original structural details compared to the conventional approach. To quantify this improvement, we define a separation metric, $sep = (I_{P1} + I_{P2}) / 2 - I_V$, where $I_{P1}$ and $I_{P2}$ are the peak intensities of the two points and $I_V$ is the intensity at the valley between them. A larger *sep* value indicates more pronounced peak separation and higher resolution. Consistent with the visual evidence, the Type 2 results yield consistently larger *sep* values, confirming the enhanced spatial resolution and fidelity provided by the end-to-end optimization.

Based on the quantitative comparisons, two primary conclusions can be drawn. First, the proposed method successfully integrates SO focusing with SI. SOSI achieves a significantly higher resolution than SO and provides a highly practical pathway to surpass the existing experimental resolution limits of conventional SO imaging systems. Furthermore, the end-to-end trained framework substantially simplifies conventional SI. Instead of relying on the empirical and subjective manual tuning of the subtraction coefficient $\gamma$, the proposed network directly targets the effective subtraction PSF by treating $\gamma$ as a learnable parameter. The optimized $\gamma$ (0.4719) differs markedly from our initial manual selection in Fig. 2 and standard values found in prior studies [21, 22]. This deviation stems from the optimized profile and amplitude matching between the jointly designed bright- and dark-field PSFs. Although a physical pinhole can suppress the massive spatial sidelobes inherent to SO, a straightforward SOSI scheme inevitably leaves residual positive and negative sidelobes at the tails of $PSF_{conf}$. In contrast, the end-to-end trained model penalizes all undesirable sidelobes outside the main lobe via the tailored loss function. This joint optimization mitigates the artifact-inducing sidelobes and further enhances the overall imaging contrast, as corroborated by the reconstructed images.

In summary, we developed an end-to-end VDINN framework for SOSI that integrates focal optimization and SI. By leveraging SI, SO focusing achieves a reduced synthetic FWHM, surpassing conventional limits while maintaining energy efficiency. Given its flexibility, the proposed framework is readily extensible to diverse wavefront engineering and detection strategies, offering a versatile pathway for high-performance super-resolution imaging.

**Funding.** National Natural Science Foundation of China (12304434, 12004362, 12204446). Fundamental Research Funds for the Provincial Universities of Zhejiang (2024YW10, 2023YW01).

**Acknowledgments.** The authors would like to thank the AI training platform for supporting this work provided by High-Flyer AI. (Hangzhou High-Flyer AI Fundamental Research Co., Ltd.).

**Disclosures.** The authors declare no conflicts of interest.

**Data availability.** Data underlying the results presented in this paper are not publicly available at this time but may be obtained from the authors upon reasonable request.